# The Collector, the Glitcher, and the Denkbilder:
## Towards a Critical Aesthetic Theory of Video Games

Jan Cao
East China Normal University

Abstract: To examine the aesthetics of video games, this paper proposes to consider video games as a contemporary multi-media version of the so-called "Denkbild," or "thought-image," an experimental genre of philosophical writing employed by members of the Frankfurt School. A poetic mode of writing, the Denkbild takes literary snapshots of philosophical, political, and cultural insights that interrupt and challenge the enigmatic form of traditional philosophical thinking. Thinking of video games through the lens of the Denkbild allows us to understand the diversity, conditionality, and incommensurability of video game as a form without reducing it to separate pieces to be examined within their respective disciplines too quickly. By presenting two snapshots of video game players, the collector and the glitcher, this paper argues that the concept of Denkbild allows us to better understand the relationships between game, gamers, and the socio-political context in terms of unexpected bonds, accidental breakthroughs, and moments of absolute freedom.

There has been much scholarly debate regarding video games as an artistic medium. Some of these hotly debated topics include: are video games art? Are they a type of mass media and, if so, what kind of experience do they offer? (e.g., Clarke and Michell, 2007; Kirkpatrick, 2011; Sharp, 2015; Atkinson and Parsayi, 2020) Such discussions show that at least some games fill the role of aesthetic experiences comparable to other forms of art, and like traditional art forms, games are played or appreciated for their own sake. But video game's interactive, often goal-oriented nature ensures a fundamentally different aesthetic experience, making it difficult to apply traditional aesthetic theories, which are often based on disinterest and aesthetic contemplation. Debates regarding the aesthetic nature of video games never seem to reach a consensus, perhaps partly due to the fact that video game as a form is often self-contradicting: they are often compared to films and television in terms of their visual presentation, but its repetitive nature and the lack of significant narrative in many games make it difficult to consider games as a meaningful medium. Games often have abstract, strict rule systems, but the philosophical concept of play is commonly defined as purposeless and carefree. Rather than further engaging with the above-mentioned debates, this paper approaches the aesthetic problem of video games from a critical theoretical perspective by answering the following questions: what role do computer games play within our contemporary aesthetics and culture? What critical potential do they have on our understanding of contemporary mass media, and how does our reception of games lead to new philosophical insights?

The theoretical structure according to which we understand video games have long been framed according to a variety of dichotomies, such as narratology and ludology, production and reception, control and agency, success and failure, to name a few. The core issue behind these debates is game scholars' ongoing effort to define what games are and their function as media objects. One of the central disputes that began in the early 2000s—the ludology and narratology debate—illustrates the contradictory and multivalent nature of video games as well as the systems of power and resistance within games and their play. The narratological perspective interprets games as a novel form of narrative that can be studied as such, whereas the ludologists assert that the study of computer games must concentrate on play, especially the kind of play that emphasizes form, structure, and goals. For the narratologists, games often emulate the look and experience of films by adopting similar cinematic techniques or introducing cut-scenes, small digital movies that are used to indicate the goals and fill the gaps in the game narrative. For example, Berys Gaut (2010: 13) claims that video game is essentially a form of digital, interactive cinema, while Aaron Meskin and Jon Robson (2010) propose to see games as a type of moving image. Others argue that video game is not necessarily a meaning-conveying medium because its creation and evolution are deeply collaborative (Lantz, 2009; Ebert, 2010). To an extent, Jesper Juul (2001) agrees by arguing that video game cannot convey ideas or represent meanings in the same way as film or novels because of its unique temporal structure, namely, there is no distinction in game play between the narrating time and the narrative time. Because the story time is now, and what comes next is not yet determined, Juul believes that it makes no sense to approach games as story-craft. As Espen Aarseth (1997: 162) indicates, "the game plays the user just as the user plays the game, and there is no message apart from the play."

Another tension within the nature of games was noted as early as around 1940 by play theorist Johan Huizinga (1950: 13), who describes play as both non-serious and absorbing: "[…] a free activity standing quite consciously outside 'ordinary' life as being 'not serious,' but at the same time absorbing the players intensely and utterly." Roger Caillois (2001: 13) further elaborates this idea by placing play-related activities on a scale from *paidia*—which refers to free and improvised play—to *ludus*, or ordered and structured game according to rules. The tension is also evident in the design of video games, because a goal-directed game is designed for ultimately purposeless play, escaping any practical identification. As Juul (2018) nicely puts it, "games make us focus on optimization and utility-seeking when playing, but also prompt us to see this optimization and utility-seeking as an experience for its own sake." The tension between purpose and purposelessness is also reflected in the socio-economic impact of the game industry. While the gaming culture is deeply shaped by capitalism—the global video game market will amount to 178 billion US dollars in market value, topping movies and music combined (Witkowski, 2021)—it also offers a platform for its own critique and resistance, including socialist video game developed by worker-owned leftist media and celebrated by the left-wing gamer community (Ottenhof, 2020), fundraising effort led by streamers, and community-building creativity in Nintendo's popular game *Animal Crossing*.

Video game is a quintessential example of hybrid, intermedia, and interactive form of art that dissolves boundaries between genres, disciplines, and modes of thinking. Rather than trying to settle the disputes between ludology and narratology, between purpose and purposelessness, this paper attempts to understand the aesthetics ramifications of video games in terms of its paradoxical character. I propose to consider video games as a twentieth-first-century multi-media version of the so-called "thought-image," or "Denkbild," an experimental genre of philosophical writing employed by members of the Frankfurt School, such as Theodor Adorno, Walter Benjamin, and Ernst Bloch. A poetic mode of writing, the Denkbild takes literary snapshots of philosophical, political, and cultural insights that interrupt the enigmatic form of traditional philosophical thinking. By constantly speaking of something beyond the dualism of the literary theory of a narrative and the philosophy of play, the Denkbild as a genre always contains its own resistance. Thinking of video games through the lens of the Denkbild allows us to understand the diversity, conditionality, and incommensurability of video game as a form without quickly reducing it to separate pieces to be examined within their respective disciplines. By presenting two snapshots of video game players, the collector and the glitcher, this paper argues that the concept of Denkbild allows us to better understand relationship between game, gamers, and the socio-political context in terms of unexpected bonds, accidental breakthroughs, and moments of absolute freedom.

## 1. The Denkbilder

The meaning of the word "Denkbild" has undergone some semantic changes since its first appearance in Stefan George's *Seventh Ring*, where he borrows the word from Dutch. George uses the word to simply mean "Idee," or idea, whereas Adorno uses it to illustrate Benjamin's writing style in *One-Way Street,* which resembles aestheticized fragments or textual snapshots of philosophical thoughts. According to Adorno, Benjamin's poetic prose "[does] not want to stop conceptual thought so much as to shock through their enigmatic form and thereby get the thought moving, because thought in its traditional conceptual form seems rigid, conventional, and outmoded" (1992: 323). One of the most significant features of the style of Denkbild is that it offers the crystallization of the origin of ideas in linguistic figurations while resisting being fully translatable into philosophical truth-claims. Benjamin's specific way of theorizing and philosophizing, according to Sigrid Weigel (1992: 50), speaks of "a third thing" beyond the dualistic opposition of poetic language and conceptual meta-discourse, of literature and philosophy. This "third thing" is the reflection, imagination and the "working-through" of a dialectical image (Weigel, 1992: 55) which cannot be neutralized or condensed into philosophical meta-discourse. An excessive, disfigured mode of writing, the Denkbild is always referring to something else, something that no longer exists. As Gerhard Richter summarizes, it is "a sign with a signifier but without a referent" (2007: 28). It thus withdrawals from transparent meaning by appearing anecdotal and interactive, full of figures and tropes that require careful deciphering. It also refuses to be contained by any institutional or political thought-system that unveils it to be transparent and self-evident.

Meanwhile, Benjamin's Denkbilder also contain a revolutionary potential. It speaks about Platonic conceptual thought by working against it, creating a kind of "intellectual short-circuiting" that strikes sparks on the familiar idea, and perhaps even "set it on fire" (Adorno, 1992: 323). A similar intellectual and aesthetic short-circuiting appears in Benjamin's analysis of the "expressionless" as a category of art that Benjamin discovers in Goethe's *Elective Affinities*, Hölderlin's hymnic poetry, and ancient Greek tragedy; as a "critical violence" against the storm of progress, the expressionless, pure language presents itself as a "caesura" that interrupts the flow of rhythm and meets the rush of ideas (2002: 354). Benjamin associate this expressionless with the essentially beautiful, which challenges the false totality of aesthetic representation. "Only the expressionless completes the work," Benjamin argues, "by shattering it into fragments of the true world and bare symbols" (2002: 340). Like the expressionless beauty of Goethe's Ottilie, Benjamin's style of the Denkbild also illuminates a philosophical idea by working against it, arresting thought in a fragmented snapshot of the true world. The revolutionary potential of the Denkbild as a work of art dwells in its capacity to deliver the shock that shatters the semblance of a coherent philosophical system of thought and resists traditional enigmatic form of thinking.

Last but not least, the Denkbild has a unique relationship with visual culture: while Adorno compare it with a "scribbled picture-puzzle," Bloch refers to it as a "snapshot," a "photomontage," a "kaleidoscope," a "considered improvisation," and a "collage of familiar objects and forgotten glances" (1979: 95-96). Among these analogies, Adorno's "picture puzzle" is particularly interesting for the purpose of this paper. A "Vexierbild," or it is commonly translated, a "hidden face," refers to an illusionistic picture that allows viewers to see images with double meanings, such as head or landscape, fruit or vase. Artists since the 16th century—Leonardo da Vinci, Giuseppe Arcimboldo, Max Ernst, Salvador Dali, to name a few—have utilized aspects of this phenomenon in both subtle and overt guises. What a "picture-puzzle" shows is not necessarily intended by its author; rather than evoking something that is expected but cannot be said, a "picture-puzzle"expects the unexpected. Like a scribbled picture-puzzle, a Denkbild expects something else alongside the intended, which is only visible to those who know where to look. This scribbled riddle echoes with Weigel's "third thing" beyond the dualistic opposition between literature and philosophy, indicates that the unexpected potentiality of the Denkbild exists in the form of a visual riddle.

If we were to resurrect this Frankfurt school experimental genre of writing with the language of today's media poetics, the contemporary Denkbild should be a form of artwork that meets the three above-mentioned criteria. It should communicate thoughts and reality while resisting any unmediated form of political intervention; embody a revolutionary potential by delivering some kind of "shock effect;" and speak about an unexpected futurity through a fundamentally visual presentation. If the Denkbild has a political function in the age of digitalized global media, it would reside in, as Richter suggests, "the invitation that it extends to us to open up the complexity of the concept of the political to its perpetual non-self-identity and thus to its unpredictable future" (2007: 23). Video games, as I argue, have the potential to complete all these tasks. As hybrid form of intermedia, video game's aesthetic potentiality dwells somewhere beyond the dualism of narratology (literature) and ludology (conceptual thought), as its deciphering always requires a "playing-through." An excessive, paradoxical, and much too playful form of contemporary form of artwork, the video-game-as-Denkbild resists any systematic philosophical interpretation, offers to illuminate some truth hidden in its puzzling character without necessarily telling the player where to look.

Thinking of video game as Denkbild helps us develop an aesthetic theory uniquely for this type of multimedia art object without the necessity to define what video game is and is not. This concept reconfigures the relationship between philosophy and art by saying what philosophy cannot say and elaborate what art does not say (see Adorno, 1997: 72). It problematizes the very notion of video game as a homogenous, self-contained genre; instead, it takes snapshots of different aspects of video game and evaluate them as signs of a larger philosophical, socio-political, and cultural semiotics. In this paper, I will look at two types of gamers, the collector and the glitcher, vis-à-vis two Benjaminian Denkbilder, "Construction Site" in the essay collection *One-Way Street,* and "The Reading Box" from his *Berlin Childhood Around the 1900s.* With the aid of contemporary media theorists Hito Steyerl and Rosa Menkman, my comparative analysis of the aesthetics of video game attempts to show readers where to look beyond the tension between meaningful narrative and meaningless form, purpose and purposelessness, *ludus* and *paidia*. While the figure of the collector presents us a network of intimate relationships among players, in-game objects, and the world around them, the glitcher promises moments of absolute freedom that interrupts the overarching narrative of progressive history and envisions a redeemed future emerged from such slivers of possibility.

## 2. The Collector

A Denkbild in Benjamin's *One-Way Street*, entitled "Construction Site," describes the joy children experience from the debris of a construction site. Rather than imitating the works of adults, children form a different, intuitive relationship with rubble and detritus. "In waste products they recognize the face that the world of things turns directly and solely to them," Benjamin writes (1979: 52-53). By playing with and collecting useless artifacts, children produce their own small world of things in the greater one. Benjamin warns those adults who wish to create things specially for children to resist the impulsion to insert their own adult activity and wishes into the things they create. Instead, they should always keep in mind the norms of this small world of artifacts, and find their way through them. What, then, are the norms of this world of things? Children are intuitively attracted by these objects because they are produced in play; the norms of this world are essentially the rules of a game. "Play," as Giorgio Agamben (2007) argues, is an entirely inappropriate use of the "sacred," or that which is removed from common use and transferred to a different sphere. Play ignores the separation between the sacred and the profane, and frees things from their sacred, appropriate use. Furthermore, by deactivating an old use, play stages the pure means or praxis that is emancipated from its relationship to an end (Agamben, 2007: 74). For example, when a cat plays with a ball of yarn as if it were a mouse, the game frees the cat's predatory activity from being necessarily directed towards the death of a mouse, rendering the activity of hunting a pure means without an end. The childish collector, by separating objects from their traditional use of the adult world, also creates a new use for things: they become the building blocks of the child's own small world.

The figure of the collector also appears in Convolute H of the *Arcades Project:*

> What is decisive in collecting is that the object is detached from all its original functions in order to enter into the closest conceivable relationship to things of the same kind. This relation is the diametric opposite of any utility, and falls into the peculiar category of completeness. What is this "completeness"? It is a grand attempt to overcome the wholly irrational character of the object's mere presence at hand through its integration into a new, expressly devised historical system: the collection. And for the true collector, every single thing in this system becomes an encyclopedia of all knowledge of the epoch, the landscape, the industry, and the owner from which it comes. (Benjamin, 1999: 204)

Here, the collector functions as a link in a chain of events and owners of the object's history. As an example, a collected book is the synthesis of two modes of history, the private memories of its collectors and an impersonal order of its catalogue. The act of collecting is therefore a form of practical memory that saves an object from being broken down into tradable pieces of data and liquidized assets. While the object is saved from capitalist ownership, it also loses its value in a mass market and becomes an empty vessel of its histories. Benjamin (1999: 209) cites the following passage from Marx twice in Convolute H, arguing that Marx offers a positive counter-type to the collector, as the collector liberates things from the drudgery of being useful: "private property has made us so stupid and inert that an object is ours only when we have it, when it exists as capital for us, or when…we *use* it." The true collector turns whatever it collects into "garbage," because the act of collecting takes objects out of the contexts of trade, ownership, and usefulness. As Kevin McLaughlin (1995: 82) points out, the collector's preservation is the opposite of commodification, as what he collects are "empty forms — mere pieces or patches that can be invested with meaning or value only because they are decontextualized." Brought out of the context of commodity and commercial exchange, the collected object loses its referent in socio-economic terms, but preserves a "chaos of memories" such as a transaction, a touch, an encounter, a city (Benjamin, 1967: 60). The pleasure of collecting lies neither in the contents nor its curation and presentation (such as in the case of a museum curator), but in the intimate relationship one forms with the object. In "The Collector," Benjamin writes, "perhaps the most deeply hidden motive of the person who collects can be described this way: he takes up the struggle against dispersion. […] by keeping in mind their affinities and their succession in time, he can eventually furnish information about his objects" (1999: 211). Collecting is in a way recreating a miniature of the universe around oneself without attempting to elucidate their meanings, properties, or relations. The collection remains an ever-ending patchwork [*Stückwerk*] that is always missing something, as the reflection of one thing is never sufficient in foreseeing the meaning of another.

Richter compares the childish collector in "Construction Site"to to Benjamin's historical materialist, the rag-picker and garbage collector of history, for he seeks his materials and inspirations in the overlooked and marginalized refuse of culture and human history, creating strategic poetic montage that reveals a revolutionary image (Richter, 2006: 135). Benjamin the historical materialist also collected his garbage of history in the reading hall of Paris's Bibliothèque Nationale, hunched over file cards, microscopic notes and old books. Despite Benjamin's Marxist argument for the collector—that he is not interested in the use or exchange value of the object, but its unique relationship with its owners and their personal history—collecting at the turn of the 20th century is not a hobby that can completely escape the curse of class oppression and capitalist modes of exchange and ownership. Born into an affluent family, Benjamin's enthusiasm for collecting partially is partly owing to the middle-class comfort and security, especially of his earlier life. In the late 19 and early 20th century, collecting things for its own sake was not the typical hobby of a lumpen proletarian, who might have never witnessed the "aura" of an authentic work of art. However, in the age of digitalized images, any ordinary working-class individual has the opportunity to collect images, especially copies of original images and films that circulate on the internet. They frequently appear in the form of memes, stickers, GIFs, screenshots, emojis that can be easily shared, edited, and used in a text or video chat.

A lot of video games—many of which are either free or could be purchase with a reasonable price—contain an artfully curated collecting experience that requires no further monetary investment. Early on in the history of video games, these collectible items are "useful" in the sense that they serve necessary functions within the game: Pac Man collected little orbs, Mario collected coins, Sonic the Hedgehog collected rings. In the recent years, more games are filled with collectible items that have no purpose in the game apart from the thrill of collecting itself. For example, each of the *The Legend of Zelda* series contains one hundred characters—Gold Skulltulas or Koroks—that can be collected. Starting from its 2007 inaugural title, the *Assassin's Creed* series encourages players to trace down a myriad of flags hidden all over the game map. In the *Witcher* series, players can collect anything from letters and cards to flowers, diagrams, books, black magic dolls, broken oar, skulls, or ashes. Most of these items, considered "junk" in the game, can be dismantled into useful crafting components or sell for money. Others (like ashes or glass) either cannot be dismantled or cannot be sold. These items are doubly useless both in the game world and in the real world, but players nevertheless collect these useless objects because "the collecting itself feels good," according to a study of Toups et al. (2016). They report that 22.8% of their participants value digital game objects due to their personal investment and memory that encapsulate these items. One participant collects Teddy Bears in the *Fallout* games because they make the game "feel like it was actually devastated by atomic war," and they collect it "as a kind of tribute to the massacred innocents" (Toups et al., 2016: 283). Another participant collects every unique item in *Skyrim*, but is particularly invested in a skull. "there is nothing really special about this skull; it doesn't have any significance and doesn't have any utility. It does have a different texture… more important there's only one of them" (Toups et al., 2016: 283). The teddy bear or skull are not "collectible" in the sense that they are outside of the game's provided schemas, unlike legendary weapons or cards. By collecting them, these players also create their own version of Benjaminian patchwork of objects that preserve affects, memories, or intimate relationships.

Another type of users collect for the sake of "social presentation" and self-expression (Toups et al., 2016: 283). As Toups et al. point out, digital game objects can be easily shared or presented automatically when players go online, and players tend to share what they have acquired and curated by recording video clips or take screenshots to share within the game platform, on social networks, and streaming websites. In the process, in-game objects are copied, reproduced, circulated, and recreated, gaining a life independent of the "original." If compared with "auratic" artworks and rare first edition books, these digital copies of images with potential copyright infringement issues have little commercial value. They are essentially the "garbage" of the digital age, uploaded and stored on the Internet, increasing their owner's "carbon footprint." The collector of in-game objects and digital waste reminds us of filmmaker and cultural critic Hito Steyerl's concept of "poor image." In her essay "In Defense of the Poor Image," Steyerl writes,

> The poor image is a copy in motion. Its quality is bad, its resolution substandard. As it accelerates, it deteriorates. The poor image is […] a lumpen proletarian in the class society of appearances, ranked and valued according to its resolution…The image is liberated from the vaults of cinemas and archives and thrust into digital uncertainty, at the expense of its own substance. The poor image tends towards abstraction: it is a visual idea in its very becoming. (2009: 1)

The wide circulation of "poor image" degrades the contemporary hierarchies established by original artworks or high-resolution images. Such hierarchies are based on the "richness" of an image in terms of its sharpness and resolution, as well as the privatization and commodification of intellectual content. These hierarchies reflect the society's conservative biases protected by "systems of national culture, capitalist studio production, the cult of mostly male genius, and the original version" (Steyerl, 2009: 3). Meanwhile, the poor image is degraded, blurry, abstract, but popular: they are made and seen by the many. As the "lumpen proletarian in the class society of appearances," poor image expresses a condition of dematerialization, shared with "capital's semiotic turn," as described by Felix Guattari: "[it]

plays in favor of the creation and dissemination of compressed and flexible data packages that can be integrated into ever-newer combinations and sequences" (1996: 202). In other words, poor images are poor because they are heavily compressed; as a result, they can travel quickly and reach a much broader audience. As images become memes, an artwork turns into an abstract idea, which is perfectly integrated in today's world of information capitalism, which thrives on previews and impression rather than immersed contemplation.

The in-game-object-as-poor-image resists the fetish value of visibility and, as a dematerialized art object, could create an alternative economy of images that circulates and reconnects dispersed players and viewers around the world, unconcerned of commercial and national agendas. The poor image constructs an alternative global network of interests, affects, and relationships, redefining the value of the image in terms of its intensity and spread. As Steyerl concludes, "By losing its visual substance it recovers some of its political punch and creates a new aura around it. This aura is no longer based on the permanence of the "original," but on the transience of the copy" (2009: 8). The new "aura" of a poor image resides in its capacity of being endlessly circulated, dispersed, recycled, and re-created, which brings about a new reality of visual art and aesthetics with its own conditions of existence. What the garbage collector of the digital age collects are no longer things, but symbols of ideas, data of memories, and translations of feelings, created by unconventional forms of circulation including pirating, torrenting, and collective editing. The circulation of these ever-changing symbols creates a network of "visual bonds" that link workers of the world with each other, creating disruptive moments of thoughts and affects that challenges today's audiovisual capitalism and the commodification of aesthetic experiences (Vertov 1995: 52).

In the world of video games, everyone becomes creators, distributors, and collectors of gamified artworks. With the revolutionary means of production and distribution of the digital age, a new portrait of the collector appears. The collector of the digital age no longer assemble antiquities, curiosities, and fragments of the past to construct their own small world, like the children on a construction site; instead, they gather sharable feelings and experiences to create an alternative economy of images, an anonymous global network of shared histories, and a collective virtual world parallel to the real one. What the collector of the digital age plays with is no longer the real thing, the originary original, but real conditions of existence or, as Steyerl (2009: 8) puts it, reality itself: " [it is] about swam circulation, digital dispersion, fractured and flexible temporalities. It is about defiance and appropriation just as it is about conformism and exploitation."

3. The Glitcher

Benjamin's B*erlin Childhood Around 1900*, an autobiographical text that offers a rich portrait of Berlin at the turn of the 20th century, describes his exploration of famous landmarks and unfamiliar interiors of the city. What distinguishes Benjamin's account from other autobiographies or travel writings around the same time is his obsession with things: a sewing box, a carousel, a telephone, a panorama… His recollection of these things become Denkbilder of the past that allowed him, in exile in 1932, to adapt to the feeling of longing homesickness. In his essay "Hope in the Past: On Walter Benjamin," Peter Szondi compares Benjamin's search for lost time with that of Proust, whose writing obviously had a strong influence on his German translator. According to Szondi, Proust searches for lost time in order to escape from time, especially future time, all-together; whereas Benjamin detects traces and hints of his later life in memories of his childhood, and in doing so, he "seeks future in the past": every place in Benjamin's memory seem to contain traces of the future (Szondi, 1978: 17-18). Citing the section "The Reading Box," Szondi argues that Benjamin's historical and social environment makes conscious reflections reflection possible. But what he does not emphasize are the openings in Benjamin's recollections: these snapshots of Denkbilder contain secret doors towards the past through which the past can be not only accessed, but also altered. Without these openings, he would not be able to "[be] sent back into the past, a past, however, which is open, not completed, and which promises the future" (Szondi, 1978: 19). These openings, as Benjamin hints in "the Reading Box," appear in the shape of letters:

> [The reading box] contained, on little tablets, the various letters…Those slender figures reposed on their slanting bed, each one perfect, and were unified in their succession through the rule of their order—the word—to which they were wedded like nuns…Yet my right hand, which sought obediently to reproduce this word, could never find the way. […] Indeed, what I seek in [the reading box] is just that: my entire childhood, as concentrated in the movement by which my hand slid the letters into the groove, where they would be arranged to form words (2006: 141).

Perhaps Benjamin means it literally when he claims that his entire childhood is concentrated in the movement of a few letters that has yet formed words. From a child's perspective, a few misspelt or misheard words could lead to a completely different world, as in the example of Benjamin's mother's sewing box. The maids tend to slur the first word as they call her "Gnädige Frau," and as a result, young Benjamin always thought they were saying "Näh-Frau," or Madam Needlework. The slur gives his mother magic that "ruled over [him] with inexorable power" (Benjamin, 2006, 113). Similarly, calling Markt-Halle "Mark-Thalle" changes everything about the market hall, from its original concept of buying and selling to the slow-moving shop owners, who become "priestesses of a venal ceres," "purveyors," "procuresses" (Benjamin, 2006: 70). Looking back at his childhood, Benjamin realizes that he could not return to that stage, when words, like games, once had magic: "My hands can still dream of this movement, but it can no longer awaken so as actually to perform it. By the same token, I can dream of the way I once learned to walk. But that doesn't help. I now know how to walk; there is no more learning to walk" (Benjamin, 2006: 141-142). What Benjamin longs is the potentiality of the word game, which allows children to freely manipulate their history and destiny like toys. For Benjamin, these word-puzzles from the past contain openings like the "narrow gate" in his *Theses on History,* through which the Messiah could enter.

The image of the reading box and its secret magic openings take on a new form in today's media: video game and its glitches. Glitches are usually slight, transient faults that can cause problems within the code and result in unexpected gaming experiences (Švelch 2014; Janik 2017). Most glitches are performed by players within the bounds of a game engine but were not intended by the game's developers. Common glitches include "clipping," which occurs when a character or object within the game environment pass through a texture or level geometry; "jumping," which indicates that the player bypasses the game's spatial limits and reaches locations that are normally not intended to be reached; "physics-based glitches," where a player utilizes a game's physics engine to perform unintended actions; and many more (Boluk and Lemieux, 2017: 15). Obviously, glitches are often exploitable by players not only to create an interesting game experience, but also to circumvent game rules, transform the game progress, and even create new grassroots game modes. Glitchers—commonly defined as players who aim to exploit weaknesses in game code and circumvent game rules and goals—form communities where they find, document, and share glitches. Glitching is often seen as a destructive practice, as it goes astray from the intended experience of a game and against the values and practices of the mainstream videogame community. Despite so, glitching also enables unintended outcomes that could potentially enrich the gaming experience. According to a player of *The Legend of Zelda* series, the beauty of this game lies in its endless potential in being broken over and again:

> " We've discovered endless glitches, tricks, sequence breaks, exploits and more leading to [The Legend of Zelda: Ocarina of Time] being beautifully broken in ways no other game is. There's a running joke among OoT [speed] runners that says anything is possible in this game. As we discover new things, this has proven to be true. We can kill enemies by idly standing near them, we can clip through walls, float on thin air, warp from the first dungeon to the last, duplicate items, obtain light arrows in the deku tree, even literally writes ones and zeros to our inventory to give ourselves new items." (Newman, 2019, 17).

For this player, *Ocarina of Time* is beautiful simply because they are free to do anything imaginable in the game world. In fact, the philosophical idea of beauty is deeply intertwined with the concept of free play. Friedrich Schiller's notion of the "play drive" frees humans from the domination of either the form drive [Formtrieb], which is concerned with dignity and principles, or the sense drive [Sachtrieb], which is concerned with self-preservation. Instead, the play drive brings them into harmony; "gives rise to freedom" (1993: 143). As play frees our minds of determinations, mankind is "in the fullest sense of the word a human being, and he is only fully a human being when he plays" (Schiller, 1993: 131) Schiller's understanding of play is not limited to a particular set of actions or objects, but can be seen as an interaction among other drives, an oscillation or back and forth movement among opposing forces without resolution. In German, the word "Spiel" contains two parallel meanings, one referring to free, purposeless play, and another to rule-based games. The design of digital games also aims at striking a balance between the protection of "play" through rules and regulations and the protection of play as "free, unfettered, and improvisational activity" (Krapp, 2011: 117). Since play is a back and forth movement, it always contains moments of breakthroughs during which absolute freedom becomes momentarily available.

Absolute freedom comes with the price of absolute chaos and potential breakdown of the entire game world; therefore, they only exist in the form of "glitches,"which originates from the German term "glitschen" (to slip, to skid). It was first used in English in 1965 to describe a spike of or change in voltage in an electrical circuit, which takes place when the circuit suddenly has a new load put on it. It has been described as a "momentary jiggle," a "disturbance," or a "fluff." By definition, as short-lived fault in a system that often corrects itself, which makes it difficult to troubleshoot, glitches are difficult to locate both spatially and temporarily. Glitchers are addicted to seeking these mysterious moment of absolute freedom in the phantom world of play. "Free Guy," an 2021 American science fiction film, visualizes the political promise of the glitcher as well as its limitations. The film tells the story of a non-player character (NPC) in a massively multiplayer online (MMO) game. The main character Guy, an NPC who lives in *Free City* and works as a bank teller, is unaware that his entire world is a virtual one within a video game. After meeting "Molotov Girl," the character of Millie Rusk, who is the developer and rightful owner of the source code of *Free City*, Guy decides to pursue her and as a result, he deviates from his programming. Taking a player's heads-up display (HUD) in the shape of a pair of sunglasses, Guy begins seeing *Free City f*rom the perspective of a player character, and progresses the game by doing good deeds (which is surprisingly not encouraged in this game). Later, viewers are told that Guy's self-awareness came from artificial intelligence code, which contains Millie's personal preferences, resulting in Guys' romantic interest in her. The two work together to obtain the source code of the prototype of *Free City* called *Life Itself*, which is a fishbowl game where players observe and peacefully interact with NPCs as the game goes on.

The term "glitch" appears in this film exactly twice. When the NPC Guy first becomes "alive," game developers refer to him as a "glitch." In *Free City*, the so-called "freedom" of a player is limited to a few things: "you can rob a store, carjack someone, punch a pedestrian in the face." As Guy decides to level up by helping people, he becomes what developers call a "glitch," as a streamer comments: "the game wasn't even made for someone like this, for a player to be a good guy. And, honestly, it's got me realizing, maybe we've been thinking about NPCs wrong this whole time." Guy quickly becomes a symbol for an alternative definition of freedom and a political leader of a digital walkout, when all the NPCs go on strike, gathering in a coffee house and refusing to perform their prescribed routine. To rally support from the community, Guy says, "Aren't you sick of being shot at? - Taken hostage? - Run over? Robbed? Stabbed? Used as a human shield?… What if we can change it?… The point is, we don't have to be spectators to our own lives. We can be whatever we want." Guy is the glitch that moves away from the established action scripts and leaps in to the world of unknown. Shocked and amazed, players around the world stop playing and log on streaming platforms to observe him and his peers, leading to a significant decline in sales of the sequence game. The glitch interrupts the ordinary experiences and discourses around the game for everyone including players, developers, and NPCs, resulting in a new way of play.

According to glitch artist Rosa Menkman's *Glitch Studies Manifesto,*

> The glitch is a wonderful experience of an interruption that shifts an object away from its ordinary form and discourse. For a moment I am shocked, lost and in awe, asking myself what this other utterance is, how was it created. Is it perhaps… a glitch? But once I named it, the momentum — the glitch — is no more…(Menkman, 2010, 5).

As Menkman argues, the glitch has no solid form or state, but denotes the generic idea of an abnormal *mode of operandi,* a moment of break from the equilibrium within a technological system, a rupture of the original experience that vanishes as soon as new conditions set in. As the glitch shocks the audience and leaves the work of art in "destructed ruins of meaning," there remains a sense of hope, "a triumphal sensation that there is something more than just devastation…a spark of creative energy that indicates that something new is about to be created" (Menkman, 2010, 5). The glitch-as-Denkbild continues to carry on the task Adorno assigned to the genre of Denkbilder, which is to shock through the traditional form of thought to get it moving again. It shatters the game world's semblance of unity into pieces of fragmented chaos, an effort that again reminds of us of Benjamin's reading of the critical violence of the expressionless and its potentiality in revealing the true work of art.

With its unique aesthetic function, the glitch shares the political task of a Marxist historian, whose work continues to haunt and interrupt the continuum of a mainstream historical narrative, usually written by the victors. By deconstructing the myth of linear progress and perfect technology, the glitch resurrects possibilities of historical change from "homogenous, empty time." In the film, the glitch guy, leader of the class struggle of repressed NPCs, offers hope by promising "a whole new world," "some kinda paradise," where the rules of *Free City* no longer applies. What he leaps into is a utopian world that appears as *Life Itself,* which is a copy of Millie and Keys' codes for their original version of the game, hidden in *Free City* in the form of an unreachable tropical island. To access this tropical island, he needs another glitch, first discovered by player and steamer Revenjamin Buttons, who claims that it has only been accessed once before the glitch was patched. Echoing with Benjamin's portrayal of the true picture of the past, which "can be seized only as an image which flashes up at the instant when it can be recognized and is never seen again" (Benjamin, 2003: 390), the gateway towards the future only exists as a glitch of the past. By visualizing the glitch that provides access to a promised utopian world, *Free Guy* (largely accidentally) shows what the Messianic narrow gate of the digital era — which allows one to seek future in the past — might look like. As Menkman points out, the glitch challenges the tendency to describe history as a continuum.

But the hope of an emancipatory utopia only exists in brief moment when the glitch happened, at least within the world of this typical Hollywood romantic comedy. The film ends with Guy admitting that he only serves as a medium, as Keys' "love letter" to Millie: "I love you, Millie. Maybe that is my programming talking, but, guess what, someone wrote that program." As the glitch becomes the new norm, player characters and NPCs enter a new capitalist equilibrium in the world of *Life Itself,* in which an alternative mode of economy that resembles the utopian fantasy of "technological democracy" overcomes the moment of revolution. Referred to as an "observational, fish bowl game," *Life Itself* is essentially a zoo-like commercial safari park where tourist-players wander around observing freely roaming NPCs. As soon as the glitch is being named as a new game, the original experience of shock "passed its momentum and vanished into a realm of new conditions" (Menkman, 2010, 5). The glitch only exists at one moment in time and cannot be preserved to a future time. As soon as it is understood as a novel form of art, the glitch would have passed its tipping point and lost its essence of its glitch-being. "If there is such a thing as technological freedom, this can only be found within the procedural momentum of glitch art, when a glitch is just about to relay a protocol," not in the creation of formally new design (Menkman, 2010, 7). By too quickly settling with the solution of a new game, the ending of "Free Guy" attempts to domesticate the glitch by containing it in a technology-operated environment, in which the glitch no longer acts to break the game or struggles against the domination of the player class. As Menkman concludes, the glitch "has now become a new commodity." Despite the commercialized Hollywood interpretation, glitchers are always looking forward to trigger the next glitch to break the equilibrium and to discover the next moment of absolute freedom.

The two Denkbilder that this paper illustrates present a new way to think about video games as aesthetically meaningful artifacts beyond the dualism of dualism of ludology and narratology, of *ludus* and *paidia*. Thinking video games as Denkbilder could also be seen as a method to deconstruct the myth of perfect technology and linear progress of history: both the collector and the glitcher offer us alternative narratives of history that analyze the flow of history as well as its ruptures and arrests. Video game is not only a medium, a narrative, a simulation, a technological thing, a toy, a sport, but also a work of art and a digital artifact in need of its own critical aesthetic theory and historiography. Thinking of video games through the concept of Denkbilder allows us to stop thinking video games either as separate pieces to be examined within different disciplines or as a piece of technology constantly progressing towards an ideal state of perfection. Video games is as diverse and exciting as it is problematic and confusing. A critical aesthetic theory about video games would challenge its established template of creative practice and reflect on its aesthetic function as a medium.